\documentclass{jfm}
\usepackage{graphicx}
\usepackage{epstopdf, epsfig}
\usepackage{xcolor}
\usepackage{amsmath}
\usepackage{bm}
\usepackage{array}
\usepackage{soul}

\shorttitle{Finite Amplitude Scaling in Transitional Pipe Flows}
\shortauthor{R. Vishnu, K. Chola}

\title{Finite Amplitude Scaling in Transitional Pipe Flows}

\author{Ravindran Vishnu \corresp{\email{1.vishnur@gmail.com}} \and Kalale Chola
}

\affiliation{Fluid Mechanics Unit, Okinawa Institute of Science and Technology Graduate University, Onna-son, Okinawa 904-0495, Japan}

\begin{document}

\maketitle

\begin{abstract}
Studies on the finite amplitude stability of pipe flows identified a range of different scaling exponents between $\beta\approx -1 $ and $\beta\approx-1.5$, relating $A\sim Re^{\beta}$, where $A$ is the minimum amplitude of disturbance to cause a transition to turbulence and $Re$ is the Reynolds number. The circumstance under which a particular scaling exponent manifests itself is still not clear. Understanding this can shed light on the different routes to turbulence \citep{willis2008experimental} and the mechanisms involved.  The exponents observed in previous experiments and simulations were explained based on the spatial localization of initial disturbances. In this paper, through direct numerical simulations (DNS), we classify the exponent, $\beta$ into two ranges; a steeper exponent with $\beta\lessapprox-1.3$ and a shallower exponent with $\beta\gtrapprox-1$. We then determine the nature of the disturbance to produce a specific exponent. 

Our results clearly show that the two ranges of the scaling exponents are related to the radial distribution of the initial disturbance, where  $\beta \lessapprox -1.3$ exists for a disturbance at the boundary, and $ \beta \gtrapprox -1$ exits otherwise. We also compare the previous experiments and simulations on injection-type and push-pull-type initial disturbances.  This study clarifies the nature of the initial disturbance that can result in either of the two different scaling exponents observed so far. 

\end{abstract}


\section{Introduction} \label{sec:intro}
One of the challenges in understanding the transition to turbulence lies in defining a critical point that determines transition, consistent across different flows.  
This critical transition point can be well defined for linearly unstable flows such as Taylor Couette and Rayleigh B\'enard convection. In contrast, the transition is complicated in linearly stable flows, such as in a pipe flow \citep{trefethen1993hydrodynamic, meseguer2003streak}, limiting the precise definition of a critical Reynolds number, $Re_c$. The transitions in pipe flows are classified as subcritical transitions since a finite amplitude disturbance is required to trigger turbulence \citep{barkley2016theoretical}. Experimentally, it has been shown that the transition depends on the nature of the disturbance and the $Re$ of the flow \citep{peixinho2007finite}. The minimal disturbance amplitude, $A$, that can trigger turbulence scales as $A\approx Re^{\beta}$ \citep{mullin2011experimental, hof2003scaling, peixinho2007finite}, where $\beta$ is the scaling exponent. Previous studies have found the range of $\beta$ to be $-1.5\lessapprox\beta\lessapprox-1$ \citep{eckhardt2007turbulence,mullin2011experimental,avila2023transition}. The exponent from experiments forced by fluid injection \citep{peixinho2007finite} were obtained numerically\citep{mellibovsky2007pipe} by an equivalent expression for a normalized amplitude \citep{schlichting2016boundary},  
 \begin{equation}
         \frac{{A}}{{A_0}}=\left(\frac{u/U_{CL}}{Re}\right)^{1/2}\bigg/\left(\frac{u/U_{CL0}}{Re_0}\right)^{1/2} ,
 \end{equation}
where $Re$ is the Reynolds number, $U_{CL}$ is the centerline velocity, $A$ is the minimal amplitude and subscript, $()_0$ refers to any reference state. Using this normalization, geometrical discrepancies between experiments and simulations were avoided \citep{mellibovsky2009critical}. By comparing with the experiments \citep{hof2003scaling}, \cite{mellibovsky2007pipe} showed that an injection-type initial condition was associated with $\beta=-1$. Further simulations using streamwise independent vortices were found to be associated with $\beta \approx-1.5$ \citep{mellibovsky2006role}, a scaling observed in experiments with the push-pull type disturbances\citep{peixinho2007finite}.

The linearly stable nature of the pipe flows demanded a nonlinear analysis to understand the transition process further. Through nonlinear optimization techniques \citep{pringle2010using,pringle2015fully,kerswell2018nonlinear}, it was shown that the optimal energy growth for transition, also called the ``minimal seed", is spatially localized. Using nonlinear optimization problem, \cite{duguet2013minimal} identified $\beta \approx 1.35$ for plane Couette flow, for the minimal disturbance. Though these studies put forth the optimal initial conditions, \cite{draad1998laminar,mellibovsky2006role} observes that realizing such simulated initial conditions in experiments may be difficult and complicated, limiting their practical applications.  

Despite the progress in understanding pipe flow transition, an important question on the nature of the initial disturbance to trigger transition is still unanswered. This pertains to the precise nature of the spatial orientation of the disturbance and its relation to the different exponents observed so far. If such a distinction in initial disturbance exists, it can shed light on how 
a distinct exponent is achieved and the path followed by the disturbance to approach a turbulent state \citep{willis2008experimental}. This is a highly relevant question, given its fundamental and practical implications for pipe flows and the need for drag reduction. Exploring this specific spatial nature of the disturbance helps to identify different flow instabilities that develop in a specific transition process and how the subsequent turbulence is initiated in pipe flows. Practically, this can decide the type of disturbance to apply for an optimal control strategy, which is relevant in drag reduction or efficient mixing scenarios. Despite its importance, the question of the precise nature of the disturbance is still open.

In this present study, we provide evidence for the nature of spatial disturbance through a series of numerical simulations initiated with the most generic, novel initial condition that can cause a transition. 
The paper is organized as follows. In section \ref{sec:numerics}, the numerical setup and the implementation of the initial disturbances are described in detail. In section \ref{sec:scaling}, we clarify the origin of the two scalings, while in section \ref{sec:compare}, we compare this with existing studies, both numerical and experiments, and summaries in section \ref{sec:conclude}.

\section{Numerical method and initial condition implementation}\label{sec:numerics}
The nondimensional 3D incompressible Navier Stokes equations, 
\begin{eqnarray}
\begin{aligned}
\partial_t\mathbf{V}+\left(\mathbf{V}\cdot \nabla\right)\mathbf{V}=-\nabla p+ { {\frac{1}{Re}}}\nabla^{2}\mathbf{V} + { {\frac{4}{Re}}}(1+\beta) \hat{z} \\
\nabla\cdot\mathbf{V}=0
\end{aligned}
 \label{eqn:MomC}
\end{eqnarray}
are solved numerically in cylindrical coordinates. The Reynolds number is defined as $Re=2UR/\nu$, $\mathbf{V}=(u_r, u_{\theta}, (1-r^2)+u_z)$ is the 3D velocity field, $U$ is the bulk velocity, $R$ is the radius of the pipe and $\nu$ is the kinematic viscosity. The length of the pipe was fixed at $L=5D$, with periodic boundary conditions at constant mass flux \citep{willis2017openpipeflow}, such that $(1+\beta)$ is the ratio of the average axial pressure gradient to the laminar axial pressure gradient. The constant mass flux helps us compare with experiments and eliminate any turbulence decay observed for constant pressure gradient cases due to flux reduction \citep{faisst2004sensitive}. A resolution of $64\times288\times288$ grid points was provided in radial, azimuthal, and axial directions. The time step was dynamically controlled using a Courant–Friedrichs–Lewy (CFL) condition. 

The short length of the pipe ($L=5D$), used in this simulation, will preclude any study of localized turbulent structures \citep{faisst2004sensitive}; for a short pipe, the turbulence typically fills the whole domain during transition \citep{willis2009turbulent} without the need to consider the advection times of the turbulent structures. Thus using a short pipe will render the critical exponent to be independent of any time window required to prescribe the initial disturbance. This time window should be a factor to be considered in long pipes \citep{mellibovsky2007pipe} and in experiments due to the transient nature of turbulence \citep{hof2008repeller, brosa1989turbulence} in pipe flows.  Moreover \cite{mellibovsky2006role} compared critical amplitude for transition for two different lengths, one longer ($L\approx50D$) and the other shorter ($L\approx10D$), and found no significant difference in the critical amplitudes. 
The resolution is confirmed from using drop-off in the amplitude of coefficients by three-four orders of magnitude in all directions. Moreover, the scaling exponents of the minimal amplitudes were compared at two grid resolutions and the maximum difference between the exponents from the two resolutions were less than 5\%, as shown in Table~\ref{tab:compare_resolution}.  Simulations presented in this study were performed for $Re$=[5300, 10000]. In the following subsection, details of the formulation of the initial disturbance are explained.

\begin{table}
  \begin{center}
\def~{\hphantom{0}}
  \begin{tabular}{llll}
Resolution & CFL & Radial location of  & \hspace{0.5cm} $\beta$  \\
 &  &  the disturbance & \hspace{0.5cm}   \\
\hline
$128\times576\times576$ & \hspace{0.13cm} 0.1 & \hspace{0.13cm} $r=0.759-1$ &  \hspace{0.5cm} $-1.57$   \\ 
$64\times288\times288$ & \hspace{0.13cm} 0.5 & \hspace{0.2cm}$r=0.761-1$ &  \hspace{0.5cm} $-1.51$  \\ 
$128\times576\times576$ & \hspace{0.13cm} 0.1 & \hspace{0.2cm}$r=0.778-0.998$ &  \hspace{0.5cm} $-1.05$    \\ 
$64\times288\times288$ & \hspace{0.13cm} 0.5 & \hspace{0.2cm}$r=0.779-0.998$ & \hspace{0.5cm} $-1.1$  \\ 
  \end{tabular}
  \caption{Comparison of scaling exponents $\beta$ obtained using different resolution}
  \label{tab:compare_resolution}
  \end{center}
\end{table}

\subsubsection{Generation of initial condition}\label{sec:ICgen}
 
The initial disturbance is introduced at a prescribed spatial location in a laminar base flow, ensuring that the disturbance is divergence-free and it does not bias the flow to any particular wave number. An ideal candidate for such a generic disturbance is a field of divergence-free random numbers. To achieve this, the disturbances comprising of uniformly distributed random numbers (between $-0.5\leq u_i\leq0.5$) for each required radial planes are generated. The disturbances are made exponentially decaying in $z$ and $\theta$ directions, by weighting the random disturbance fields, $\mathcal{R}_i$, by $c_\theta=c_z=e^{-2(z-2.5)^2}$.

In previous experiments, the disturbances were provided from the wall of the pipe by either injection or by injection-suction (also called push-pull). If we compare single orthogonal injection and injection-suction, both disturb the flow radially. However, a key difference between them arises from the induced velocities near the wall in $z$ and $\theta$ directions, due to an extra suction for the push-pull type disturbance. Thus, a minimal disturbance model required to simulate this key difference in the induced velocities is to provide the fluctuations in $u_z$ and $u_\theta$ only, keeping $u_r=0$, thus obtaining the velocity field $(u_r, u_\theta/r, u_z )= (0, c_\theta c_z \mathcal{R}_1, c_\theta c_z \mathcal{R}_2)$. This also has the advantage that divergence -free condition need to be satisfied only in the $z-\theta$ plane, as gradients of $u_r$ in radial direction is zero.

\begin{figure}
    \centering   \includegraphics[width=1\textwidth,trim={2cm 4cm 2cm 3cm},clip]{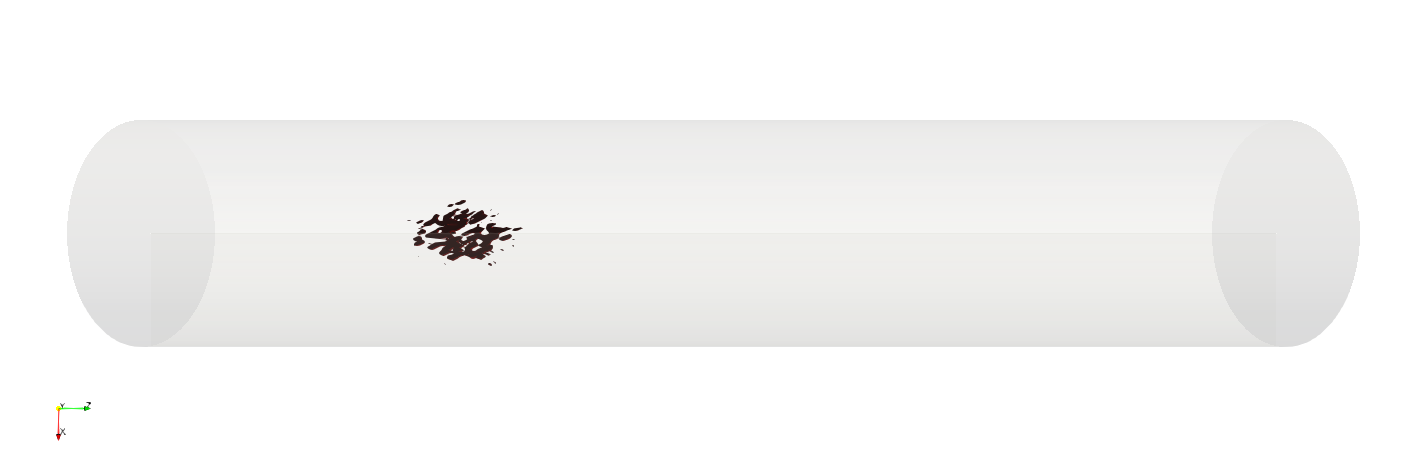}
    \caption{Illustration of the initial disturbance shown by the contour of the velocity magnitude ($\sqrt{u_\theta^2+u_r^2+u_z^2}$, a value of 0.01 shown here). Here, the initial disturbance is exponentially decaying in all three directions and is superimposed onto  the base laminar flow. }
    \label{fig:IC_contour}
\end{figure}

The disturbances ($u_z$ and $u_\theta/r $ in the $z-\theta$ plane) are orthogonalized relative to the wavenumber vector using the Gram-Schmidt method in Fourier space, ensuring the resulting velocity field in physical space is divergence-free. 
The energy spectrum of this resulting vector field is scaled to equal energies for all wave numbers, ensuring the most generic disturbance without biasing to a particular wavenumber. Moreover, a 2/3 dealiasing rule is applied for a nonlinear operation in the Fourier space. Finally, an inverse fast Fourier transform is performed to recover the divergence-free disturbance velocities in the physical space. The generated divergence-free fields in $z-\theta$ planes are appended to the base laminar velocity field for the required radial planes and are used as the initial condition for each simulation. The variations in the random initial conditions are limited to only variations in its amplitude \citep{faisst2004sensitive}, scaling them by a factor, $\mathcal{A}$, $i.e.$ ($\mathcal{A} u_\theta, \mathcal{A} u_z$). Each simulation then evolves from these initial disturbances of different amplitudes and the minimum amplitude that triggers a transition is investigated by varying $ \mathcal{A} $. An illustration of the initial disturbance is shown in Fig~\ref{fig:IC_contour}. The figure shows the velocity fluctuations in the $r-z$ plane at $t=0$, where the initial disturbance is exponentially decaying in all three directions. Next we discuss on quantifying the transition process.

\subsection{Quantifying the transition}\label{sec:quantify}
\begin{figure}
    \centering   \includegraphics[scale=0.35,trim={0cm 0cm 0cm 0cm},clip]{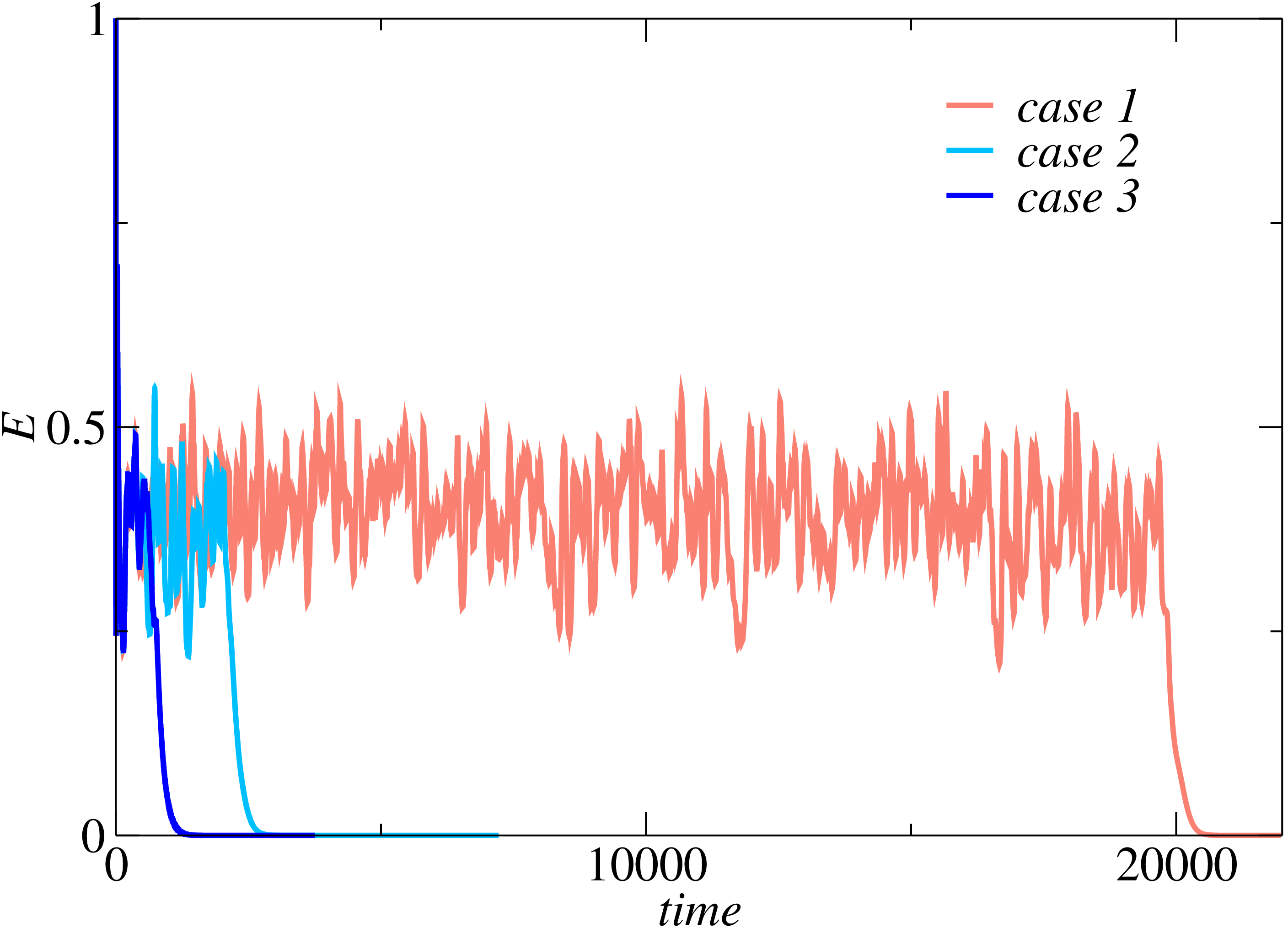}
    \caption{The evolution of volumetric kinetic energy shows that the turbulence is transient and can relaminarise randomly. This corresponds to three cases at the same $Re=2530$, initiated with the same initial condition.}
    \label{fig:trans_turb}
\end{figure}

In the present study, the evolution of total volumetric kinetic energy, $E$ of perturbation to the mean flow, is chosen as an indicator of the transition.   
Figure \ref{fig:trans_turb} shows its evolution for three identical simulations, performed with the same initial condition at $Re=2530$. The initial disturbance amplitude is above the minimum amplitude required to trigger turbulence; however, the time at which the system relaminarises, $t_{lam}$ is different in each case, i.e., at $t_{lam}\approx 1000$ (case 3), $2000$ (case 2), and $20000$ (case 1). This confirms the probabilistic nature of the relaminarization process \citep{brosa1989turbulence,faisst2004sensitive,darbyshire1995transition}.
\begin{figure}
    \centering   \includegraphics[scale=0.35,trim={0.0cm 0.0cm 0.0cm 0.0cm},clip]{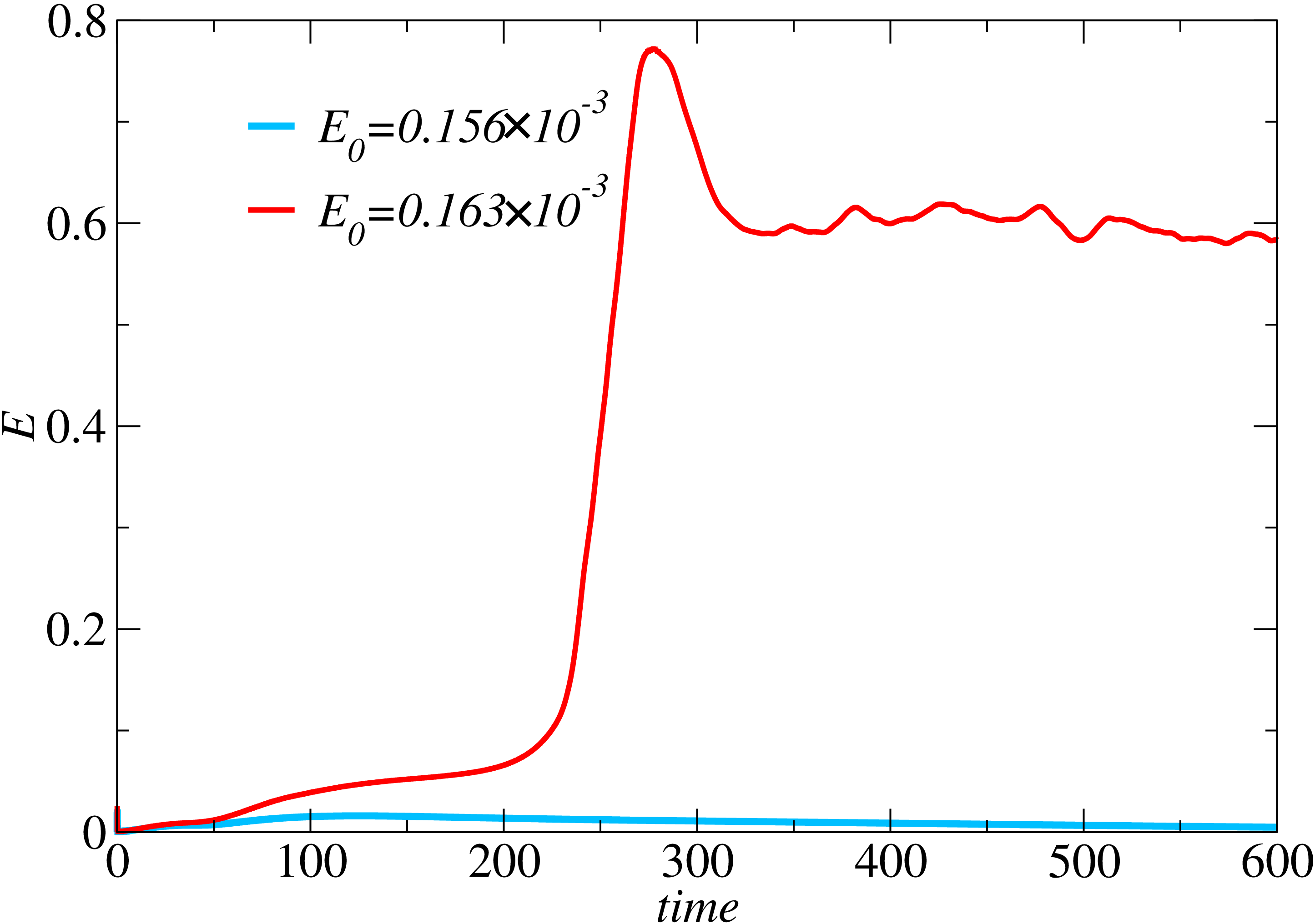}
    \caption{The evolution of volumetric kinetic energy, $E$ for $Re=6000$ shows that transition is hindered when the initial energy is low (blue curve) and is triggered with the minimum initial disturbance (red curve).}
    \label{fig:trans_turbLam}
\end{figure}

 Turbulence in pipes is considered a transient phenomenon \citep{hof2008repeller}, and any event of an impending relaminarization happens after a turbulent transition, requiring the need to quantify a transition using only the growth to the turbulent state without considering the later stage of relaminarization.  Moreover, as mentioned by \cite{faisst2004sensitive,o1994transient}, the trajectories settle to the turbulent state within about $150$-time units, well within our time span of growth to the turbulent state, sufficient to quantify it as a transition. Here, we focus only on the transition process by examining the initial growth or decay of  $E$ without considering the impending relaminarization of the turbulence. 
 Thus, a statistical analysis based on large ensembles, also computationally expensive, is not undertaken, and the transition was determined solely based on the initial response of $E$. 
As an example, the evolution of $E$ for $Re = 6000$ is shown in the figure~\ref{fig:trans_turbLam} for two similar disturbance levels, one higher and other lower than the threshold. If the disturbance amplitude is enough to trigger a transition (red curve), it is considered a transition event, while initial disturbances (blue curve) not strong enough cause a decay event. 
 
In the figure, a $5\%$ higher amplitude than the decaying disturbance (blue curve) causes a growth (red curve) and we choose it as a transition event.

\section{Scaling of the disturbance amplitude}\label{sec:scaling}
 Several experiments and numerical simulations investigated the scaling of finite amplitude disturbance in pipe flows \citep{hof2003scaling,peixinho2007finite, mellibovsky2006role,mellibovsky2007pipe,avila2023transition}. The disturbance is usually implemented experimentally through an injection type \citep{hof2003scaling} or an injection-suction type mechanism \citep{peixinho2007finite}. In the former injection type, a needle (or a combination of needles) injects the disturbance fluid orthogonally or tangentially to disturb the flow. In the latter case (also called push-pull type), there is an additional needle near the point of injection to simultaneously remove an equal quantity of fluid by suction, ensuring that the total mass flux of the disturbance is zero. In experiments with the injection type disturbance, \cite{peixinho2007finite} and \cite{hof2003scaling} obtained a scaling exponent near to $-1$ while \cite{peixinho2007finite} obtained an exponent of $-1.3$ and $-1.5$ using a push-pull disturbances. Similar scaling exponents were also reported by \cite{mellibovsky2006role,mellibovsky2009critical,meseguer2003streak} in numerical studies. A review of the results based on the disturbance type, $Re$ range studied, and the corresponding scaling exponent, $\beta$, is summarised in Table~\ref{tab:compare_Beta}.

\begin{table}
  \begin{center}
\def~{\hphantom{0}}
  \begin{tabular}{lll}
$Re$ & $\beta$ & Disturbance type \& Source \\
\hline
$\approx$ 3000-30000 & -2/3 & Exp: Small $k$ periodic disturbance  \citep{draad1998laminar}  \\ 
2000-18000 & -1$\pm$0.01 & Exp: Six holes tangential injection  \citep{hof2003scaling}  \\ 
$\approx$ 2000-10000 & -1 & Exp: Single orthogonal jet  \citep{peixinho2007finite}  \\ 
$\approx$ 2000-10000 & -1 & Exp: Six tangential jets  \citep{peixinho2007finite}  \\ 
$\approx$ 3000-30000 & -1 & Exp: Large $k$ suction \& blowing separated by $180^o$  \citep{draad1998laminar}\\ 
$\approx$ 3500-11000 & -1 & Sim: Six holes tangential injection  \citep{mellibovsky2009critical}  \\ 
$\approx$ 5000-12600 & $\approx$ -1.06 & Sim: Azi. $k$=3, with random 3D noise  \citep{mellibovsky2006role}  \\ 
$\approx$ 5000-12600 & $\approx$ -1.10$\pm$0.03 & Sim: Azi. $k$=2, with random 3D noise \citep{mellibovsky2006role}  \\ 
$\approx$ 2000-10000 & -1.3$\pm$0.1 & Exp: Streamwise/spanwise push-pull  \citep{peixinho2007finite}  \\ 
$\approx$ 5000-12600 & $\approx$ -1.35$\pm$0.02 & Sim: Azi. $k$=1, with random 3D noise \citep{mellibovsky2006role}  \\ 
$\approx$ 2500-12600 & $\approx$ -1.47$\pm$0.02 & Sim: Azi. $k$=1, with optimal 3D noise \citep{mellibovsky2006role}  \\ 
$\approx$ 2000-10000 & -1.5$\pm$0.1 & Exp: Oblique/Anti-oblique push-pull \citep{peixinho2007finite}  \\

$\approx$ 5000-10000 & -1.5 & Sim: Streamwise $k$=1.5 \citep{meseguer2003streak}  \\ 
  \end{tabular}
  \caption{Comparison of scaling exponents $\beta$ obtained using different types of disturbances implemented in experiments (Exp) and numerical simulations (Sim), where Azi.=azimuthal; $k$=wavenumber}
  \label{tab:compare_Beta}
  \end{center}
\end{table}
It is clear from Table~\ref{tab:compare_Beta} that two distinct scalings are observed, where $\beta \approx -1$ is associated with injection type disturbances and $\beta \lessapprox -1.3$ is associated with push-pull disturbances. It can be inferred from \cite{peixinho2007finite, darbyshire1995transition} that a localized disturbance can be obtained using the push-pull disturbance, and using such a disturbance in oblique direction, the transition to turbulence was through streaks and hairpin vortices \citep{peixinho2007finite}, while the injection-type disturbance gave a catastrophic transition \citep{mellibovsky2007pipe}. However, in all these studies, the definition of localization was not precise with respect to the direction in which localization should be imposed to obtain the specific scaling exponents. 

\begin{figure}
    \centering   \includegraphics[width=1\textwidth,trim={0cm 0cm 0cm 0cm},clip]{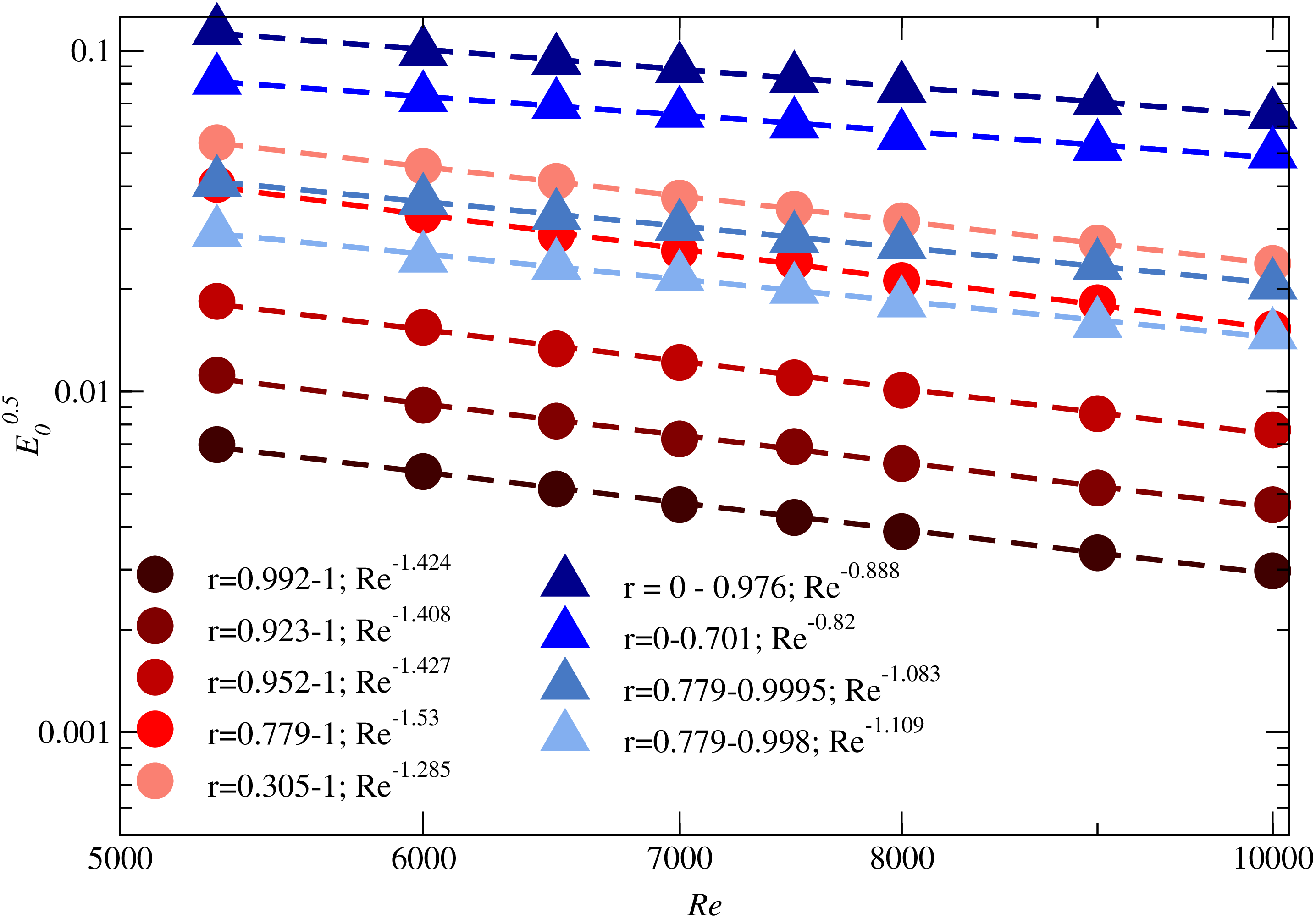}
    \caption{The minimal amplitude of disturbance for different $Re$ scales differently depending on spatial distribution of the disturbance in the radial direction. A disturbance originating at the wall has a higher slope of $\beta \lessapprox -1.3$, while those avoiding the wall have a shallower slope of $\beta \approx-1$.}
    \label{fig:IC_sqrtKE}
\end{figure}

We explore the precise nature of localization of the disturbance to invoke different scaling exponents in figure \ref{fig:IC_sqrtKE}.
In the figure, the minimal disturbance to trigger the transition, quantified by $\sqrt{E_0}$, (where $E_0$ is the energy, $E$ at $t=0$), is plotted against $Re$ for different forms of the initial disturbances.

The red curves (circles) are cases with initial disturbances originating at the wall and extending inwards in the radial direction. Meanwhile, the blue curves (triangles) are initial disturbances initiated not at the wall and extending inwards in the radial direction.  From the figure, it is clear that all the initial disturbances except those originating at the wall have a lower slope, with exponent $\beta\approx -1$. However, when initial disturbances are initiated at the wall, they have a higher slope with $\beta\lessapprox-1.3$.
Moreover, the figure shows that disturbances originating at the wall trigger turbulence with lesser energies than disturbances originating not at the wall, indicating that the wall disturbances are also the most optimal disturbances. This is consistent with the findings of \cite{duguet2013minimal}, where the minimal disturbances for plane Couette flow, obtained using nonlinear optimization gave a higher slope of $\beta =-1.35$. 

Thus the change in the radial distribution of the initial disturbance indicates an associated change in the scaling exponent $\beta$. Particularly, all the disturbances, except those originating at the wall, have $\beta\approx -1$, while those originating at the wall, extending inwards, have $\beta\lessapprox -1.3$. To obtain this higher slope, we suggest that initial disturbance originates at the wall and be exponentially decaying inward in the radial direction, and to obtain a $\beta \approx -1$, the disturbance grows from 0 at the wall radially inwards. These are suitable initial disturbances for an experimental implementation. To further support this claim, in the next section, we outline the results from earlier studies and compare with our results.

\section{Comparison with existing results} \label{sec:compare}
The earlier experimental works related to radial disturbances can be traced back to the works by \cite{draad1998laminar, durst2006forced} and \cite{peixinho2007finite}. In the work by  \cite{draad1998laminar}, they provided push-pull disturbance at $180^o$ separation between them, creating an effect similar to injection alone due to the considerable separation between injection and suction, and exhibits an exponent $-1$ for long wave number cases. \cite{peixinho2007finite} provided the disturbance by push-pull at proximity, and thus obtained an exponent $\beta\approx-1.5$, a result which we obtain with axial - azimuthal fluctuations at the wall. 
The present results also clarify the inference from earlier literature \citep{mellibovsky2007pipe} that though an injection-type disturbance is localized, it exhibits an exponent $\beta \approx -1$, similar to the results we obtain when perturbations are not at wall.

The numerical simulations by \cite{meseguer2003streak, mellibovsky2006role} initialized with streamwise vortices with low azimuthal modes,  gave $\beta \approx -1.5$.  However,  studies on such optimal modes and vortical structures that give maximum transient growth are difficult to implement experimentally \citep{mellibovsky2006role, draad1998laminar}. This underlines the advantage of the present study in suggesting an experimentally realizable disturbance, which is optimal. A striking similarity in the exponent observed for disturbances at the wall, that can trigger turbulence with minimal initial amplitudes for pipe flows and optimal minimal disturbance for plane Couette flow \citep{duguet2013minimal} can be observed. Thus, from our study, we can infer that a  disturbance originating at the wall can provide an optimal disturbance and higher exponent as observed in push-pull experiments.

Our present study clarifies the notion of optimal disturbance to include wall disturbances also and connects the gap between the appropriate spatial distribution of the initial disturbance and its associated exponent observed in the scaling amplitude.

\section{Conclusion}\label{sec:conclude}

This paper provides clear evidence that the spatial distribution of the initial disturbance in a pipe flow causes different transition scenarios. This is exhibited in the scaling exponent $\beta$, observed with different disturbances in many previous studies. This study highlights that the transition to turbulence in pipe flow differs depending on whether it is initiated by perturbing the wall or elsewhere. Thus, the claim by \cite{darbyshire1995transition} that the transition process is insensitive to the form of disturbance seems questionable, as the transition scenario differs depending on the initial disturbances.  We posit that the push-pull disturbance creates fluctuation at the wall to generate an exponent $\beta\lessapprox-1.3$. Moreover, disturbances originating at the wall are also the most efficient disturbances to trigger turbulence. This study thus rekindles the interest in the question of the origin behind the existence of the two distinct scaling exponents during the transition to turbulence in pipe flows. 
Since the initial disturbances evolve with different exponents into turbulence, this classification of initial disturbances may explain how the turbulence is developed in different scenarios, opening avenues for studying different routes to turbulence.

We acknowledge Pinaki Chakraborty (OIST), and Lin Li (Sichuan University), for constructive discussions throughout this research. We also thank Ashley Willis (University of Sheffield) for making OPENPIPEFLOW \citep{willis2017openpipeflow} openly available. This work was supported by the Okinawa Institute of Science and Technology Graduate University (OIST), and the simulations were performed at the Diego High-Performance Computing facility at OIST.\\ \\
\textbf{Funding.} This work was supported by the Okinawa Institute of Science and Technology Graduate University.\\ \\
\textbf{Declaration of Interests.} The authors report no conflict of interest.

\bibliographystyle{jfm}

\begin{thebibliography}{26}
\expandafter\ifx\csname natexlab\endcsname\relax\def\natexlab#1{#1}\fi
\def\au#1{#1} \def\ed#1{#1} \def\yr#1{#1}\def\at#1{#1}\def\jt#1{\textit{#1}} \def\bt#1{#1}\def\bvol#1{\textbf{#1}} \def\vol#1{#1} \def\pg#1{#1} \def\publ#1{#1}\def\arxiv#1{#1}\def\org#1{#1}\def\st#1{\textit{#1}}

\bibitem[Avila {\em et~al.\/}(2023)Avila, Barkley \& Hof]{avila2023transition}
{\sc \au{Avila, Marc}, \au{Barkley, Dwight} \& \au{Hof, Bj{\"o}rn}} \yr{2023}  \at{Transition to turbulence in pipe flow}.  \jt{Annual Review of Fluid Mechanics}  \bvol{55},  \pg{575--602}.

\bibitem[Barkley(2016)]{barkley2016theoretical}
{\sc \au{Barkley, Dwight}} \yr{2016}  \at{Theoretical perspective on the route to turbulence in a pipe}.  \jt{Journal of Fluid Mechanics}  \bvol{803},  \pg{P1}.

\bibitem[Brosa(1989)]{brosa1989turbulence}
{\sc \au{Brosa, Ulrich}} \yr{1989}  \at{Turbulence without strange attractor}.  \jt{Journal of statistical physics}  \bvol{55},  \pg{1303--1312}.

\bibitem[Darbyshire \& Mullin(1995)]{darbyshire1995transition}
{\sc \au{Darbyshire, AG} \& \au{Mullin, T}} \yr{1995}  \at{Transition to turbulence in constant-mass-flux pipe flow}.  \jt{Journal of Fluid Mechanics}  \bvol{289},  \pg{83--114}.

\bibitem[Draad {\em et~al.\/}(1998)Draad, Kuiken \& Nieuwstadt]{draad1998laminar}
{\sc \au{Draad, Adrianus~Antonius}, \au{Kuiken, GDC} \& \au{Nieuwstadt, FTM}} \yr{1998}  \at{Laminar--turbulent transition in pipe flow for newtonian and non-newtonian fluids}.  \jt{Journal of Fluid Mechanics}  \bvol{377},  \pg{267--312}.

\bibitem[Duguet {\em et~al.\/}(2013)Duguet, Monokrousos, Brandt \& Henningson]{duguet2013minimal}
{\sc \au{Duguet, Yohann}, \au{Monokrousos, Antonios}, \au{Brandt, Luca} \& \au{Henningson, Dan~S}} \yr{2013}  \at{Minimal transition thresholds in plane couette flow}.  \jt{Physics of Fluids}  \bvol{25}~(8).

\bibitem[Durst \& {\"U}nsal(2006)]{durst2006forced}
{\sc \au{Durst, Franz} \& \au{{\"U}nsal, B{\"u}lent}} \yr{2006}  \at{Forced laminar-to-turbulent transition of pipe flows}.  \jt{Journal of Fluid Mechanics}  \bvol{560},  \pg{449--464}.

\bibitem[Eckhardt {\em et~al.\/}(2007)Eckhardt, Schneider, Hof \& Westerweel]{eckhardt2007turbulence}
{\sc \au{Eckhardt, Bruno}, \au{Schneider, Tobias~M}, \au{Hof, Bjorn} \& \au{Westerweel, Jerry}} \yr{2007}  \at{Turbulence transition in pipe flow}.  \jt{Annu. Rev. Fluid Mech.}  \bvol{39},  \pg{447--468}.

\bibitem[Faisst \& Eckhardt(2004)]{faisst2004sensitive}
{\sc \au{Faisst, Holger} \& \au{Eckhardt, Bruno}} \yr{2004}  \at{Sensitive dependence on initial conditions in transition to turbulence in pipe flow}.  \jt{Journal of Fluid Mechanics}  \bvol{504},  \pg{343--352}.

\bibitem[Hof {\em et~al.\/}(2008)Hof, De~Lozar, Kuik \& Westerweel]{hof2008repeller}
{\sc \au{Hof, Bj{\"o}rn}, \au{De~Lozar, Alberto}, \au{Kuik, Dirk~Jan} \& \au{Westerweel, Jerry}} \yr{2008}  \at{Repeller or attractor? selecting the dynamical model for the onset of turbulence in pipe flow}.  \jt{Physical review letters}  \bvol{101}~(21),  \pg{214501}.

\bibitem[Hof {\em et~al.\/}(2003)Hof, Juel \& Mullin]{hof2003scaling}
{\sc \au{Hof, Bj{\"o}rn}, \au{Juel, Anne} \& \au{Mullin, T}} \yr{2003}  \at{Scaling of the turbulence transition threshold in a pipe}.  \jt{Physical review letters}  \bvol{91}~(24),  \pg{244502}.

\bibitem[Kerswell(2018)]{kerswell2018nonlinear}
{\sc \au{Kerswell, RR}} \yr{2018}  \at{Nonlinear nonmodal stability theory}.  \jt{Annual Review of Fluid Mechanics}  \bvol{50},  \pg{319--345}.

\bibitem[Mellibovsky \& Meseguer(2006)]{mellibovsky2006role}
{\sc \au{Mellibovsky, Fernando} \& \au{Meseguer, Alvaro}} \yr{2006}  \at{The role of streamwise perturbations in pipe flow transition}.  \jt{Physics of Fluids}  \bvol{18}~(7).

\bibitem[Mellibovsky \& Meseguer(2007)]{mellibovsky2007pipe}
{\sc \au{Mellibovsky, Fernando} \& \au{Meseguer, Alvaro}} \yr{2007}  \at{Pipe flow transition threshold following localized impulsive perturbations}.  \jt{Physics of Fluids}  \bvol{19}~(4).

\bibitem[Mellibovsky \& Meseguer(2009)]{mellibovsky2009critical}
{\sc \au{Mellibovsky, Fernando} \& \au{Meseguer, Alvaro}} \yr{2009}  \at{Critical threshold in pipe flow transition}.  \jt{Philosophical Transactions of the Royal Society A: Mathematical, Physical and Engineering Sciences}  \bvol{367}~(1888),  \pg{545--560}.

\bibitem[Meseguer(2003)]{meseguer2003streak}
{\sc \au{Meseguer, Alvaro}} \yr{2003}  \at{Streak breakdown instability in pipe poiseuille flow}.  \jt{Physics of Fluids}  \bvol{15}~(5),  \pg{1203--1213}.

\bibitem[Mullin(2011)]{mullin2011experimental}
{\sc \au{Mullin, Thomas}} \yr{2011}  \at{Experimental studies of transition to turbulence in a pipe}.  \jt{Annual Review of Fluid Mechanics}  \bvol{43},  \pg{1--24}.

\bibitem[O’sullivan \& Breuer(1994)]{o1994transient}
{\sc \au{O’sullivan, PL} \& \au{Breuer, KS}} \yr{1994}  \at{Transient growth in circular pipe flow. ii. nonlinear development}.  \jt{Physics of Fluids}  \bvol{6}~(11),  \pg{3652--3664}.

\bibitem[Peixinho \& Mullin(2007)]{peixinho2007finite}
{\sc \au{Peixinho, Jorge} \& \au{Mullin, Tom}} \yr{2007}  \at{Finite-amplitude thresholds for transition in pipe flow}.  \jt{Journal of Fluid Mechanics}  \bvol{582},  \pg{169--178}.

\bibitem[Pringle \& Kerswell(2010)]{pringle2010using}
{\sc \au{Pringle, Chris~CT} \& \au{Kerswell, Rich~R}} \yr{2010}  \at{Using nonlinear transient growth to construct the minimal seed for shear flow turbulence}.  \jt{Physical review letters}  \bvol{105}~(15),  \pg{154502}.

\bibitem[Pringle {\em et~al.\/}(2015)Pringle, Willis \& Kerswell]{pringle2015fully}
{\sc \au{Pringle, Chris~CT}, \au{Willis, Ashley~P} \& \au{Kerswell, Rich~R}} \yr{2015}  \at{Fully localised nonlinear energy growth optimals in pipe flow}.  \jt{Physics of Fluids}  \bvol{27}~(6).

\bibitem[Schlichting \& Gersten(2016)]{schlichting2016boundary}
{\sc \au{Schlichting, Hermann} \& \au{Gersten, Klaus}} \yr{2016} {\em Boundary-layer theory\/}.  \publ{springer}.

\bibitem[Trefethen {\em et~al.\/}(1993)Trefethen, Trefethen, Reddy \& Driscoll]{trefethen1993hydrodynamic}
{\sc \au{Trefethen, Lloyd~N}, \au{Trefethen, Anne~E}, \au{Reddy, Satish~C} \& \au{Driscoll, Tobin~A}} \yr{1993}  \at{Hydrodynamic stability without eigenvalues}.  \jt{Science}  \bvol{261}~(5121),  \pg{578--584}.

\bibitem[Willis {\em et~al.\/}(2008)Willis, Peixinho, Kerswell \& Mullin]{willis2008experimental}
{\sc \au{Willis, AP}, \au{Peixinho, Jorge}, \au{Kerswell, RR} \& \au{Mullin, T}} \yr{2008}  \at{Experimental and theoretical progress in pipe flow transition}.  \jt{Philosophical Transactions of the Royal Society A: Mathematical, Physical and Engineering Sciences}  \bvol{366}~(1876),  \pg{2671--2684}.

\bibitem[Willis(2017)]{willis2017openpipeflow}
{\sc \au{Willis, Ashley~P}} \yr{2017}  \at{The openpipeflow navier--stokes solver}.  \jt{SoftwareX}  \bvol{6},  \pg{124--127}.

\bibitem[Willis \& Kerswell(2009)]{willis2009turbulent}
{\sc \au{Willis, Ashley~P} \& \au{Kerswell, Rich~R}} \yr{2009}  \at{Turbulent dynamics of pipe flow captured in a reduced model: puff relaminarization and localized ‘edge’states}.  \jt{Journal of fluid mechanics}  \bvol{619},  \pg{213--233}.

\end{thebibliography}

\end{document}